\documentclass{iopjournal}

\usepackage{amssymb}
\usepackage{amsmath}

\newcommand{\Y}[1]{{\color{black} {\bf} #1}}

\begin{document}

\articletype{Paper} %	 e.g. Paper, Letter, Topical Review...

\title{Inverse design of exceptional points in a single-resonance two-port network}

\author{Y.F. Li$^1$, Biao Chen$^{2,*}$\orcid{0000-0001-6311-9306}, Y. Wu$^{3, 4}$, Y. Liu$^{1,*}$\orcid{0000-0002-8814-8573}, H. Lin$^4$ and Bin Zhou$^{1, 5}$}

\affil{$^1$Department of Physics, School of Physics, Hubei University, Wuhan, 430062, P. R. China}

\affil{$^2$Department of Applied Physics, Aalto University, Espoo, 02130, Finland}

\affil{$^3$School of Electronic Science and Engineering, University of Electronic Science and Technology of China, Chengdu, 611731, P. R. China}

\affil{$^4$College of Physics Science and Technology, Central China Normal University, Wuhan, 430079, P. R. China}

\affil{$^5$Wuhan Institute of Quantum Technology, Wuhan, 430206, P. R. China}

\affil{$^*$Author to whom any correspondence should be addressed.}

\email{biao.chen@aalto.fi, yangjie@hubu.edu.cn}


%%Research highlights
%\begin{highlights}
%\item Propose an analytical method to inversely find the exceptional point of a scattering matrix for a two-port network. 
%\item Use equivalent circuit model to verify its feasibility. 
%\end{highlights}

\begin{abstract}
Exceptional points (EPs) in non-Hermitian photonic systems enable unconventional control of wave amplitude and phase. However, identifying the EPs in a multidimensional parameter space of a system can be nontrivial and, in some cases, even infeasible. Here we propose an inverse-design method to efficiently locate the scattering EPs for a two-port resonant system supporting a single mode. The proposed method provides a direct way for tuning of geometric parameters to realize scattering EPs, as confirmed by both full-wave simulation and equivalent circuit model. In principle, our method is compatible with multi-mode system and applicable to a broad class of resonant systems. 
\end{abstract}

\section{Introduction}
\label{sec1}
Exceptional points (EPs) are non-Hermitian (NH) spectral singularities at which both eigenvalues and eigenvectors merge of an open system \cite{Heiss2012, Miri2019,Zhang2026}. They have emerged as a unifying concept in a broad range of wave platforms, including optics \cite{Ruter2010, Feng2014}, microwave \cite{Dietz2011}, acoustics \cite{Shen2018}, electronics \cite{Choi2018} and cold-atom systems \cite{Pan2019}. In photonics, EPs are closely associated to parity-time (PT) symmetry and its generalizations, and have been shown to underlie a variety of unconventional effects, such as asymmetric mode switching \cite{Doppler2016,Ghosh2016}, topological mode transfer by dynamically encircling an EP \cite{Xu2016}, and abrupt phase transitions between different scattering phases~\cite{Chong2011}. Recent work has also pushed EPs physics beyond simple two-mode systems, leading to higher-order EPs \cite{Hodaei2017}, exceptional line~\cite{He2025} and even exceptional surfaces such as bulk Fermi arcs and Weyl exceptional rings \cite{Zhou2018}, as well as time-modulated and nonlinear platforms where EPs strongly affect the dynamics of light \cite{Arkhipov2024}. These advances highlight the EPs as a powerful candidate for the control of resonances, dispersion and wave transport in NH photonic structures. 

%\comments{[From you statement, I do not think they are different 'approaches' as you said, but just different platforms (two-port system) supporting scattering EPs. Please correct this description and also give citations for different platforms]}.

%\section*{CRediT authorship contribution statement}
%\textbf{Y.F. Li:}Writing-original draft, Software, Investigation, Data curation. \textbf{B. Chen:}  \textbf{Y. Wu:} \textbf{Y. Liu:}  \textbf{H. Lin:} \textbf{Bin Zhou:}

Despite rapid progress in the realization of scattering EP, most methods on different platforms typically involve coupled resonators \cite{Zhu2018}, PT-symmetric cavities \cite{Feng2014}, and metasurfaces with engineered loss and performing extensive parameter sweeps \cite{Dong2020}. However, these designs are often tied to specific topologies and tuning parameters (e.g., gain-loss contrast, coupling strength, or unit-cell geometry), making it difficult to locate EP in general. From a theoretical perspective, the scattering matrix of a single-resonant structure near resonance is primarily determined by a small set of effective parameters, such as the resonance frequency, quality factor, and port phases \cite{Joannopoulos2008,Christopoulos2024}. On the other hand, a practical inverse-design method that induce EP conditions in the complex S-parameters to experimentally accessible quantities in generic resonant networks is still lacking. 

In this work, we propose an inverse-design method for EP in single-resonant two-port systems. As a prototype, we consider a  square-ring shaped resistive film sandwiched between two dielectric layers, and the incident wave propagates along $z$ axis, forming a scattering system \cite{Chen2022}. The NH nature of the system comes from the lossy open modes, while the resistive film introduces material loss and tunes the mode Q factor, leading to complex eigenfrequencies. We use a linear approximation to obtain an analytical frequency-dependent scattering matrix, $\mathbf{S}(f)$, in the vicinity of the resonance frequency $f_{\mathrm{res}}$. Based on $\mathbf{S}(f)$, the EP condition and the corresponding EP frequency ($f_{\mathrm{EP}}$) can be derived analytically. However, since $\mathbf{S}(f)$ only captures the dispersion near the resonance, the EP condition is physically satisfied only when $f_{\mathrm{EP}}$ coincides with $f_{\mathrm{res}}$. Practically, by tuning the sheet resistance of the film and the thicknesses of the dielectric layer, $f_{\mathrm{EP}}$ and $f_{\mathrm{res}}$ are shifted towards each other and an EP is realized when the two frequencies coincide. The trajectories of the scattering eigenvalues of frequency exhibit clear coalescence on resonance, providing direct evidence of an EP in the system~\cite{Dembowski2001}. 

To gain an intuitive interpretation of this model, we construct an equivalent circuit model (ECM) \cite{Yu2025}. This ECM consists of a series RLC resonance embedded between two transmission lines \cite{Wu2024}. Compared with this previous work, the advantage of our model is clearer exceptional dispersion near EP. We obtain scattering eigenvalue trajectories and EP signatures that are excellent fitting with those from the full-wave simulation. The good correspondence between the two methods confirms that our method captures the essential resonant and phase mechanisms responsible for the EP.

Overall, our results demonstrate that, for systems dominated by a single resonance, the important parameters for realizing an EP are the resonance Q-factor and the port phase, which can be independently controlled by the internal loss of the resonance and the thickness of cladding layers. The proposed method provides an analytical and experimentally accessible route to inversely design EPs in two-port system.

\section{Modeling and calculation}

\begin{figure}[b!]
	\centering
	\includegraphics[width=.6\columnwidth]{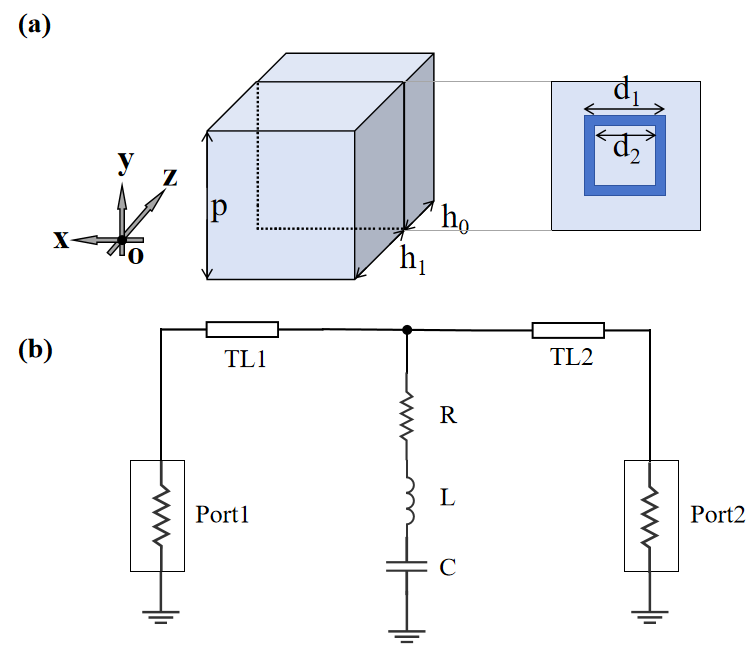}
	\caption{(a) Specific structure of the simulation model, the two dielectric layers both have length and width $d$, with thicknesses $h_0$ and $h_1$, respectively, and a square ring resistive film is sandwiched between them. Its outer and inner radius are $d_1$ and $d_2$, respectively. (b) The equivalent circuit of the simulation model in which port 1 and port 2 correspond to the air regions, TL1 and TL2 correspond to the dielectric layers, and the series RLC network correspond to the resistive film in the model. }
	\label{FIG:1}
\end{figure}

We consider a multilayered planar structure composed of a resistive film of square ring shape sandwiched between two dielectric layers, as the unit cell shown in Fig.~\ref{FIG:1}(a). The thicknesses of cladding layers are respectively $h_{0}$ and $h_{1}$ with the same permittivity $\varepsilon$. The resistive film is formed as an annular ring with outer radius $d_{1}$, inner radius $d_{2}$ and sheet resistance $R_{s}$. The outer radius $d_1$ and inner radius $d_2$ jointly determine the effective area and conductive path length of the resistive square ring film, which determines the intrinsic inductance and capacitance of the series RLC equivalent circuit. Furthermore, the thickness of the substrate also changes the resonance frequency in a negative-correlated manner. Periodic boundary conditions are imposed in the transverse directions, while input and output ports are placed in the air regions on both sides of the multilayer to define a two-port scattering situation.

We construct an ECM of this structure, as shown in Fig.~\ref{FIG:1} (b). In this model, port 1 and port 2 are assigned to the input and output of the circuit respectively and the impedance of both ports is chosen as $Z_{0} =377\Omega$, equal to the impedance of free space. Two transmission lines (TL1 and TL2)  correspond to the dielectric layers, the series RLC network correspond to the resistive film in the model, and are therefore represented by identical transmission-line with characteristic impedance $Z_{1}$ and propagation constant $\beta$. They differ only in their physical thicknesses $h_{0}$ and $h_{1}$. The resistive ring film is modeled by a series RLC branch with resistance $R$, inductance $L$, and capacitance $C$, which is connected in parallel between the two transmission-line in the circuit. The overall ECM thus consists of a two-port network formed by the first transmission-line section (dielectric layer 1), the series RLC branch (resistive film), and the second transmission-line section (dielectric layer 2), both of its two ports are air regions.

From simulation we obtain the complex $2\times2$ scattering matrix $S(f)$ at each frequency for different parameter settings in the model. The frequency range is fixed between 5 and 15 GHz, $f_{\rm{res}}$ is defined as the frequency at which the magnitude of the transmission coefficient $\vert S_{21}(f)\vert$ reaches minimum. To find $f_{\mathrm{EP}}$ for the system, we use an analytic method in the vicinity of $f_{\mathrm{res}}$. Each element $S_{ij}(f)$  is approximated by a linear function of frequency, which yields an analytic $2\times2$ matrix $S(f)$ and expressions for its eigenvalues $\lambda_{1,2}(f)$. For a two-port system the eigenvalues can be written as $\lambda_{1,2}(f)=[\mathrm{tr}S(f)\pm\sqrt{\Delta(f)}]/2$, where $\mathrm{tr}$ denotes the trace of $S(f)$, and $\Delta(f)$ denotes the discriminant which depends on $S_{11}$, $S_{12}$, $S_{21}$ and $S_{22}$, i.e. $\Delta(f)=(S_{11}-S_{22})^2+4S_{12}S_{21}$, whose detailed derivation can be found in Appendix \ref{Ana.}. In this way, for each set of structure parameters, we can obtain both $f_{\mathrm{res}}$ and $f_{\mathrm{EP}}$. Since the construction of the EP is based on the resonance of the system, our design strategy is to tune the parameters until these two frequencies coincide, $f_{\mathrm{res}}\approx f_{\mathrm{EP}}$. When this coincidence is achieved, the scattering eigenvalues $\lambda_{1,2}(f)$ exhibit coalescence. Based on the associated eigenvalue spectral trajectory, we can distinguish an EP, at which both the eigenvalues and eigenvectors coalesce, from an ordinary eigenvalue degeneracy point at which only the eigenvalues are degenerate.

\begin{figure}[t!]
	\centering
	\includegraphics[width=0.9\columnwidth]{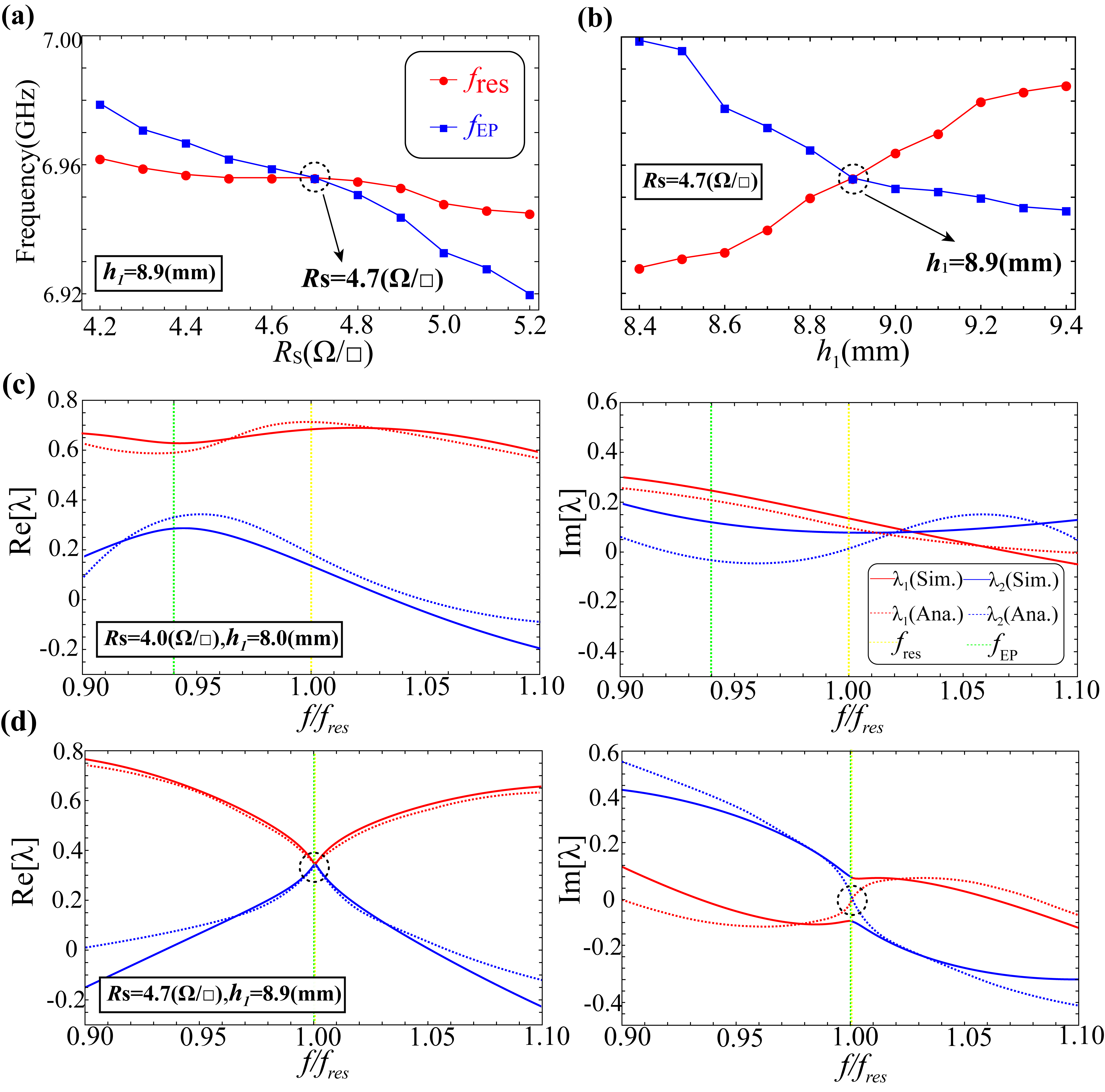}
	\caption{The variations of the two frequencies when (a) $R_s$ and (b) $h_1$ are individually varied with the other parameters fixed, the red curve represents $f_{\mathrm{res}}$ and the blue curve represents $f_{\mathrm{EP}}$. (c) and (d) show the eigenvalue plots under different conditions of the analytical and simulated solutions which (c) $R_s=4.0~\Omega/\square$, $h_1=8.0~\mathrm{mm}$ and (d) $R_s=4.7~\Omega/\square$, $h_1=8.9~\mathrm{mm}$. The red and blue curves represent the two eigenvalues$(\lambda_1,\lambda_2)$, while the solid and dashed lines represent the simulated and analytical solutions, and the yellow and green dashed lines indicate $f_\mathrm{res}$ and $f_\mathrm{EP}$.}
	\label{FIG:2}
\end{figure}

\section{Parameter tuning}

To reach the EP condition, $f_{\mathrm{res}}\approx f_{\mathrm{EP}} = 6.965 \mathrm{GHz}$, we examine the dependence of these frequencies on the geometric parameters. We first manually choose a set of geometry parameters for the FSS operating in microwave band, as  ${p}_{1}=10~\mathrm{mm}$, $p_{2}=8~\mathrm{mm}$, $d=15~\mathrm{mm}$, $h_{0}=6~\mathrm{mm}$. The sheet resistance $R_{s}$ and the thicknesses of the top dielectric layer $h_{1}$ are the parameters related to the quality factor of resonance and port phase, respectively. By tuning $R_{s}$ and $h_{1}$, $f_{\mathrm{res}}$ and  $f_{\mathrm{EP}}$ are shifted with different behaviors, which makes their coincidence possible. The parameters sweep results on $R_s$ and $h_{1}$ are shown in Fig.~\ref{FIG:2}(a) and (b), respectively. Furthermore, such influences of $R_{s}$ and $h_{1}$ on $f_{\mathrm{res}}$ and $f_{\mathrm{EP}}$ can be also understood from the ECM, where $R_s$ determines the resistance $R$ in series RLC branch and $h_1$ determines the length of the first transmission-line ($\mathrm{TL1}$), i.e. the port phase.

 As shown in Fig.~\ref{FIG:2}(a) and (b),  $f_{\mathrm{res}}$ and $f_{\mathrm{EP}}$ coincide at $6.96\mathrm{GHz}$, when $R_s=4.7\, \Omega/\square$ and $h_1=8.9\, \mathrm{mm}$, at which point the EP condition is satisfied. 
We present the eigenvalues of scattering matrix [cf. detailed derivation in \ref{Ana.}] of two devices in Fig.~\ref{FIG:2}(c-d), where solid and dash lines are the results from full-wave simulation and ECM. For comparison, the frequency axes are normalized by their resonance frequencies. The green and yellow vertical dash lines indicate $f_{\mathrm{EP}}$ and $f_{\mathrm{res}}$, respectively. When $f_{\mathrm{EP}}$ deviates from $f_{\mathrm{res}}$, the system does not exhibit the scattering EP [see example in Fig.~\ref{FIG:2}(c)], and if they coincide, the exceptional point for the two eigenvalues occurs at the resonance frequency as shown in Fig.~\ref{FIG:2}(d). It is noted that a small deviation in right panel (d) originates from the deviation of weak nonlinear frequency dispersion inherent to the simulated resonant structure from the first-order linear approximation model, which may lead to numerical error in its practical realization which could be further considered in our future work. This minor discrepancy does not undermine the EP behavior observed in our work, but means a small deviation of wave-nature simulation off from a simple two by two scattering matrix as shown in Fig.~\ref{FIG:3}(a) inset at resonance frequency slightly off from our EP.

\section{S-parameter fitting and results}
The wave propagation in the system can be described by the scattering matrix $\mathbf{S}$. It can be expressed as:
\begin{eqnarray}
\begin{bmatrix}
 b_1\\b_2
\end{bmatrix}=\mathbf{S}\begin{bmatrix}
 a_1\\a_2
\end{bmatrix},
\end{eqnarray}
\begin{eqnarray}\label{2}
\mathbf{S}=\begin{bmatrix}
 S_{11}&S_{12}\\
 S_{21}&S_{22}
\end{bmatrix}=\begin{bmatrix}
 R_\mathrm{f}&T_\mathrm{f}\\
 T_\mathrm{b}&R_\mathrm{b}
\end{bmatrix}.
\end{eqnarray}
In this work, we use the transfer matrix method (TMM) to get S-parameters. TMM provides a convenient way to analyze the scattering responses of the cascading system. The total transfer matrix of the proposed system is expressed as:
\begin{eqnarray}\label{3}
\begin{bmatrix}
 V_\mathrm{1}\\I_\mathrm{1}
\end{bmatrix}=\mathbf{T}_\mathrm{total}\begin{bmatrix}
 V_\mathrm{2}\\I_\mathrm{2}
\end{bmatrix},\quad\mathbf{T}_\mathrm{total}=\begin{bmatrix}
 A &B \\
  C&D
\end{bmatrix}.
\end{eqnarray}
$V_\mathrm{1,2}$ and $I_\mathrm{1,2}$ represents the voltage and current at port1 and port2 of the network, respectively. For a cascaded network, the total ABCD matrix is obtained by multiplying the individual matrices in the order of propagation. In our case, the top and bottom dielectric layers are represented by transmission-lines with given characteristic impedance $Z_{1}$ and propagation constant $\beta$ with their lengths of $h_0$ and $h_1$, respectively. The transfer matrices are written as:
\begin{eqnarray}\label{4}
\mathbf{T}_\mathrm{TL1}=\begin{bmatrix}
 \cos (\beta h_1)&iZ_1\sin (\beta h_1) \\
{i\sin(\beta h_1) }/{Z_1}  &\cos(\beta h_1 ) 
\end{bmatrix},
\end{eqnarray}
\begin{eqnarray}\label{5}
\mathbf{T}_\mathrm{TL2}=\begin{bmatrix}
 \cos (\beta h_0)&iZ_1\sin (\beta h_0) \\
{i\sin(\beta h_0) }/{Z_1}  &\cos(\beta h_0 ) 
\end{bmatrix},
\end{eqnarray}
where $\beta={2\pi f\sqrt{\epsilon}}/{c_0}$, $\epsilon=2.2$, $c_0\approx3\times 10^8\mathrm{m/s}$, $ Z_1={Z_0}/{\sqrt{\epsilon}}\approx 254\Omega$, $Z_1$ denotes the characteristic impedance of the transmission-line. The resistive square ring film layer is represented by a series RLC resonance circuit, as shown in Fig. \ref{FIG:1}(b). The corresponding transfer matrix is:
\begin{eqnarray}\label{6}
\mathbf{T}_\mathrm{RLC=}\begin{bmatrix}
 1 &0 \\
Z_\mathrm{f}^{-1} &1
\end{bmatrix},
\end{eqnarray}
where $Z_\mathrm{f}=I\omega L_1+{1}/{I\omega C_1}+R_1$, $\omega=2\pi f$. The total ABCD matrix of the proposed single resonance device can be expressed as:
\begin{eqnarray}\label{7}
\mathbf{T}_\mathrm{total}&=&\mathbf{T}_\mathrm{TL1} \mathbf{T}_\mathrm{RLC}\cdot\mathbf{T}_\mathrm{TL2}\\
&=&
\begin{bmatrix}
 \cos (\beta h_1)&iZ_1\sin (\beta h_1) \\
{i\sin(\beta h_1) }/{Z_1}  &\cos(\beta h_1 ) 
\end{bmatrix}
\begin{bmatrix}
 1 &0 \\
Z_{f}^{-1}   &1
\end{bmatrix}
\begin{bmatrix}
\cos (\beta h_0)&iZ_1\sin (\beta h_0) \\
{i\sin(\beta h_0) }/{Z_1}  &\cos(\beta h_0 ) 
\end{bmatrix}\\
&=&\begin{bmatrix}
 A&B \\
 C &D 
\end{bmatrix}.
\end{eqnarray}
According to the transformation relation between the transfer matrix and scattering matrix, the reflection and transmission coefficients can be expressed as:
\begin{eqnarray}\label{8}
S_{11}=\frac{AZ_1+B-CZ_1^2-DZ_1}{AZ_1+B+CZ_1^2+DZ_1},
\end{eqnarray}
\begin{eqnarray}\label{9}
    S_{12}=\frac{2Z_1}{AZ_1+B+CZ_1^2+DZ_1}.
\end{eqnarray}
By the calculation of Eqs.~\eqref{8} and~\eqref{9}, we can obtain the reflection coefficient $S_{11}(R_\mathrm{f})$ and transmission coefficient $S_{12}(T_\mathrm{f})$ when the wave is incident from forward. When the wave is incident from backward, according to Eq.~\eqref{3}, the following equation can be obtained:
\begin{eqnarray}\label{10}
\begin{bmatrix}
 V_\mathrm{b}\\I_\mathrm{b}
\end{bmatrix}=\begin{bmatrix}
    A&B\\
    C&D
\end{bmatrix}^{-1}\begin{bmatrix}
 V_\mathrm{f}\\I_\mathrm{f}
\end{bmatrix}.
\end{eqnarray}
Similarly, when the wave is incident from port $2$, the reflection coefficient $S_{22}(R_\mathrm{b})$ and the transmission coefficient $S_{21}(T_\mathrm{b})$ can be expressed as:
\begin{eqnarray}\label{11}
S_{22}=\frac{-AZ_1+B-CZ_1^2+DZ_1}{AZ_1+B+CZ_1^2+DZ_1},
\end{eqnarray}
\begin{eqnarray}\label{12}
S_{21}=\frac{2\det(\mathbf{T}_\mathrm{total})Z_1}{AZ_1+B+CZ_1^2+DZ_1}.
\end{eqnarray}
At this point, we obtained the scattering matrix $\mathbf{S}$ of the designed multilayer planar structure.

\begin{figure}[t!]
	\centering
	\includegraphics[width=.5\columnwidth]{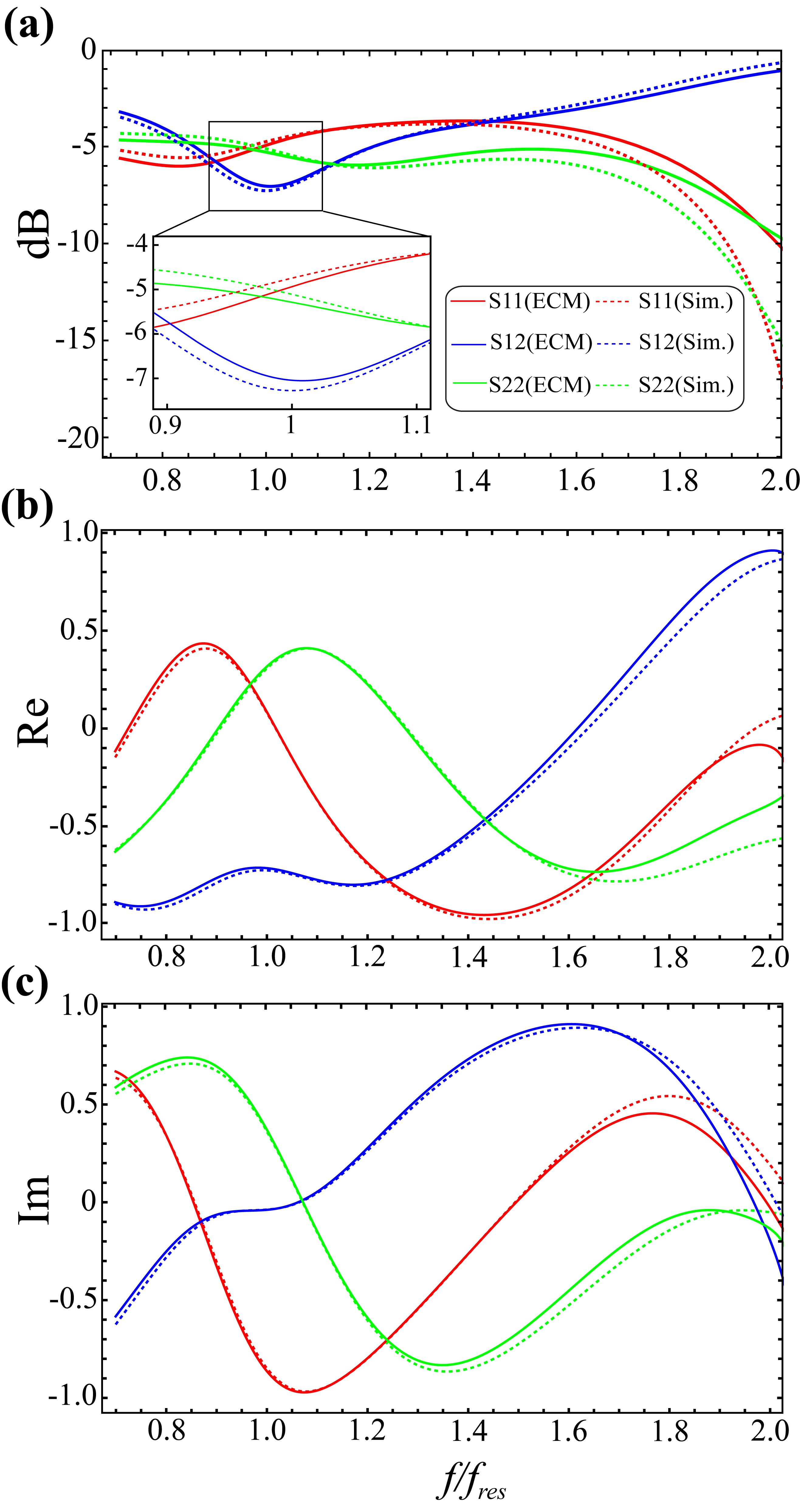}
	\caption{(a) Comparison of the fitted S-parameter magnitudes obtained by the ECM and simulation, the solid and dashed lines correspond to the ECM and simulation, the red, blue, and green curves represent$S_{11},S_{12},S_{22}$, respectively. Comparison of (b) real part and (c) imaginary part obtained by the two methods.}
	\label{FIG:3}
\end{figure}

We determine the circuit parameters $R$, $L$, and $C$ by curve fitting the S-matrix obtained from ECM to that from full-wave simulations. The constraints of the ECM are imposed using these S-parameters as the reference. We selecting several frequencies, then the frequency in the ECM is fixed, and the solution of its S-matrix is exactly these complex values, which turns the problem into solving multiple equations involving only the RLC parameters. Note that the complex S-matrix is fitted for both real and imaginary parts. Finally, we obtain a set of parameters, $R=76.48\Omega, L=5.85\mathrm{nH}, C=0.0891\mathrm{pF}$, that gives the same S-matrix as the simulation. Fig.~\ref{FIG:3} (a-c) shows the comparison for the magnitude, real part, and imaginary part of S-parameters between ECM and full-wave simulations. It should be clarified that the above ECM is only used for post-hoc validation rather than inverse design in this study. The high consistency of $S$-parameters further validates the equivalent circuit’s physical reliability. For complex multilayered structures, our analytical EP-solving method can be directly applied on the fitted ECM to avoid extensive full-wave scanning, effectively lowering computational costs for EP parameter searching. We note that our work reveals reflection higher than transmission in both forward and backward incidences, instead of perfect backward transmission in \cite{Wu2024}.

\begin{figure}[t!]
	\centering
	\includegraphics[width=.5\linewidth]{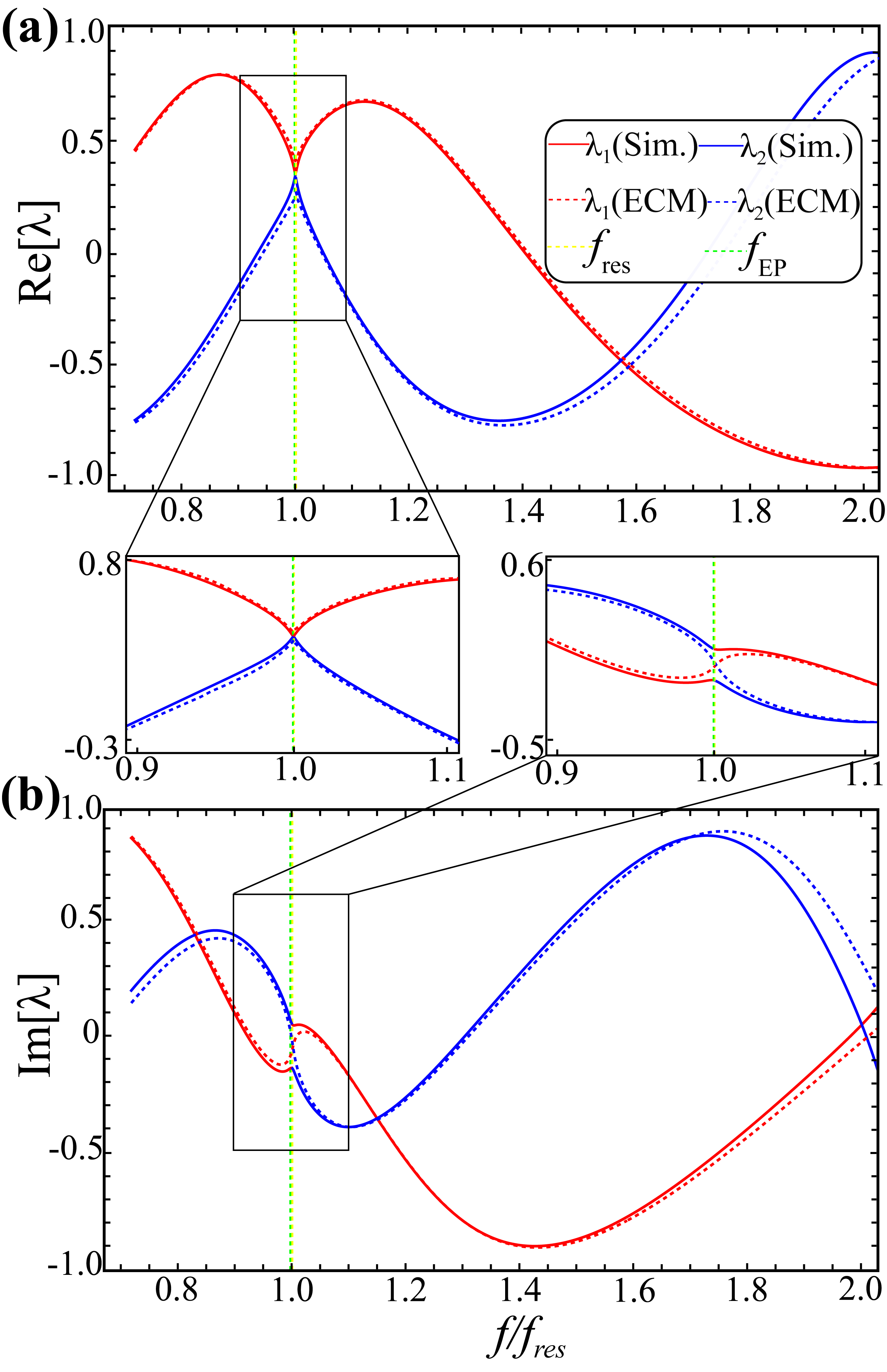}
	\caption{Comparison of the eigenvalue plots of the simulation and ECM over the operating frequency range. (a) real part (b) imaginary part, and their enlarged plots in the vicinity of $f_\mathrm{res}$. The red and blue curves represent the two eigenvalues$(\lambda_1,\lambda_2)$, while the solid and dashed lines correspond to the simulation and ECM. }
	\label{FIG:4}
\end{figure}

Obviously, ECM and simulation show a good fitting. To further investigate the characteristics of the EP, it is necessary to plot the eigenvalue diagrams of both methods and compare them. For a scattering matrix $\mathbf{S}$, its eigenvalue $\lambda$ can be numerically solved from $\det(\mathbf{S}-\lambda\mathbf{I})=0$, i.e.
\begin{eqnarray}\label{13}
    \lambda_{1,2}=\frac{1}{2}\bigg[(S_{11}+S_{22}) \pm\sqrt{(S_{11}+S_{22})^2-4(S_{11}S_{22}-S_{12}S_{21})}\bigg].
\end{eqnarray}
Then we calculate eigenvalue shown in Fig. \ref{FIG:4}. It can be observed that over the entire frequency band, the fitting between the simulation and ECM is good. The zoomed-in insets for the eigenvalue trajectories in the middle of Fig. \ref{FIG:4} demonstrate the proposed two-port system exhibits an exceptional point at the resonance frequency.

\section{Conclusion}
In this work, we present an inverse design method for realizing EP in a planar multilayer structure composed of a resistive ring film sandwiched between two dielectric layers, and ECM is constructed corresponding to structure. By varying only two adjustable parameters $R_s$ and $h_1$, one can bring convergent the two frequencies, i.e. $f_{\mathrm{res}}= f_{\mathrm{EP}}$, and the eigenvalue curves exhibit the EP. Furthermore, the analytical framework of our method relies on linearization near an independent resonant frequency to locate corresponding EP frequencies. Therefore it is not limited on the single-mode feature of the original scattering system, and\Y{could well apply for \emph{multi-mode} structures under weak intermode coupling}. To do so, we can separate each resonant mode via the transfer matrix method to construct independent sub-scattering matrices for each, then apply the same linear approximation and EP frequency solving workflow one by one, and tune its loss or phase parameters to match frequency separately. Overall, our results demonstrate that, for a single-resonant two-port structure, the key parameters for engineering an EP are the internal loss and the port phase. \Y{We note that our method sidesteps the heavy computation loads in navigating parameter spaces}, and analytically inverse design via making use of linearized scattering matrix method to find out the EP point in one shot. Henceforth this proposed method provides an analytical and experimentally-implementable way to realize scattering EPs.

\section*{Declaration of competing interest}
The authors are not aware of any affiliations, memberships, funding, or financial holdings that might be perceived as affecting the objectivity of this work.

\bibliographystyle{iopart-num}
\bibliography{LiY2025refs3}
% \ack{Sample text inserted for demonstration.}

\funding{We are supported by Natural National Science Foundation of China (Grant Nos. 62571212, 11804087, U25D8012), Science and Technology Department of Hubei Province (2024AFA038, 2022CFB553, 2022CFA012), Program of Outstanding Young and Middle-aged Scientific and Technological Innovation Team of Colleges and Universities in Hubei Province (T2020001), the Wuhan City Key R\&D program (2025050602030069), and 2023 supplemental grant for 1A0702E004 Modern Optics from Graduate School of Hubei University. }
% This section is a list of funder names and grant numbers

\roles{\textbf{Y.F.~Li}: Data curation, Formal analysis, Investigation, Software, Validation, Visualization, Writing -- original draft, Writing -- review \& editing. 
\textbf{B.~Chen}: Conceptualization, Formal analysis, Investigation, Methodology, Software, Validation, Writing -- review \& editing. 
\textbf{Y.~Wu}: Conceptualization, Methodology. 
\textbf{Y.~Liu}: Conceptualization, Formal analysis, Funding acquisition, Investigation, Project administration, Supervision, Writing -- review \& editing. 
\textbf{H.~Lin}: Funding acquisition, Resources. 
\textbf{B.~Zhou}: Funding acquisition, Supervision, Writing -- review \& editing. }

\data{The data that support the findings of this study are available within the article.} 
% For more information on IOP Publishing's research data policy see: https://publishingsupport.iopscience.iop.org/questions/research-data/

%\suppdata{ }

\newpage
\appendix
\section{Analytical solution derivation}
\label{Ana.}
To obtain an analytic expression and plot its eigenvalues, we proceed as follows. 

First, we choose a fixed $f_\mathrm{res}$ (the minimum of $\left|S_{21}\right|$), all subsequent approximations are calculated in a small frequency range around $f_\mathrm{res}$(e.g. $f_\mathrm{res}\pm0.2\mathrm{GHz}$), according to Eq.~\eqref{2}, $\mathbf{S}$ is a $2\times2$ complex matrix:
\begin{eqnarray}
    \mathbf{S}(f)=\begin{bmatrix}
        S_{11}(f)&S_{12}(f)\\
        S_{21}(f)&S_{22}(f)
    \end{bmatrix},
\end{eqnarray}
is approximated by a first-order polynomial in frequency:
\begin{eqnarray}
    S(f)\approx S(f_0)+S'(f_0)(f-f_0),
\end{eqnarray}
In the following, we approximate $S(f)$ by its linear expansion around $f_\mathrm{res}$, and for simplicity, write as $  S(f)=S(f_0)+S'(f_0)(f-f_0)$. The $2\times2$ matrix has four elements, and each element has two coefficients, so in total we obtain eight complex coefficients. 

Second, within the narrow linear valid range around $f_\mathrm{res}$, the linearized scattering matrix can be used to determine the $f_\mathrm{EP}$. An EP occurs when the two eigenvalues and their corresponding eigenvectors fully coalesce of the $2\times2$ scattering matrix.

We set:
\begin{eqnarray}
x = f - f_\mathrm{res}.
\end{eqnarray}

\Y{For a $2\times2$ complex matrix and Eq.~\eqref{13}, such coalescence condition is equivalent to:
\begin{eqnarray}\label{condition}
\big(\mathrm{tr}\,\mathbf{S}(f)\big)^2 = 4\det \mathbf{S}(f), 
\end{eqnarray}
which expands as
\begin{eqnarray}
\big(S_{11}+S_{22}\big)^2 = 4\big(S_{11}S_{22}-S_{12}S_{21}\big). 
\end{eqnarray}
We note that Eq.~\eqref{condition} may appear superficial by the simple counterexample $\mathbf{S}(f)=\lambda \mathbf{I}$, which satisfies it and contains no EP. Thus to confirm the presence of EP one needs to examine the rank of $(\mathbf{S}-\lambda \mathbf{I})$ so that the matrix is truly defective and hence contains an EP.   }

The first‑order Taylor linearization of $\mathbf{S}(f)$ is only mathematically valid within a narrow frequency neighborhood centered at $f_{\rm res}$. Extrapolating this linear model far away from $f_{\rm res}$ will introduce errors caused by higher‑order nonlinear dispersion of the scattering response. Accordingly, the solved EP frequency is physically meaningful only under the self‑consistent constraint $f_{\rm EP}=f_{\rm res}$. Any frequency solution outside the linear valid domain is non‑physical and excluded in our analysis. We set:
\begin{eqnarray}
    S_{11}(x)=S_{11}(x_0)+S'_{11}(x_0)x,\\
     S_{12}(x)=S_{12}(x_0)+S'_{12}(x_0)x,\\
      S_{21}(x)=S_{21}(x_0)+S'_{21}(x_0)x,\\
       S_{22}(x)=S_{22}(x_0)+S'_{22}(x_0)x; 
\end{eqnarray}

Substituting the linearly approximated S‑parameters into this condition, we can analytically solve for the $f_\mathrm{EP}$. Since the linear S‑matrix approximation is only physically reliable in the immediate vicinity of $f_\mathrm{res}$, only one self‑consistent physical solution exists within this narrow frequency window. 

Note that we distinguish an exceptional point (EP) when both the eigenvalues and eigenvectors coalesce, but for ordinary degeneracy point, only the eigenvalues are degenerate by examining the eigenvalue trajectories. In particular, the two eigenvalue branches exhibit the characteristic branch exchange across the EP, as shown by the exchange of their real parts in Fig. 2(d).

\end{document}